\documentclass[preprint]{ptephy}

\preprintnumber{KYUSHU-HET-149}

\usepackage{amssymb}
\usepackage{amsthm}
\usepackage{amsmath}
\usepackage{booktabs}
\usepackage{mathbbol}
\DeclareMathOperator{\tr}{tr}

\newcommand{\Slash}[1]{{\ooalign{\hfil/\hfil\crcr$#1$}}}
\numberwithin{equation}{section}

\usepackage{url}
\let\Gamma\varGamma
\let\Delta\varDelta
\let\Theta\varTheta
\let\Lambda\varLambda
\let\Xi\varXi
\let\Pi\varPi
\let\Sigma\varSigma
\let\Upsilon\varUpsilon
\let\Phi\varPhi
\let\Psi\varPsi

\begin{document}

\title{Universal formula for the flavor non-singlet axial-vector current from
the gradient flow}

\author{%
\name{\fname{Tasuku} \surname{Endo}}{1},
\name{\fname{Kenji} \surname{Hieda}}{2},
\name{\fname{Daiki} \surname{Miura}}{3},
and
\name{\fname{Hiroshi} \surname{Suzuki}}{4,\ast}
}

\address{%
\affil{1,2,3,4}{Department of Physics, Kyushu University, 6-10-1 Hakozaki, Higashi-ku, Fukuoka, 812-8581, Japan}
\email{hsuzuki@phys.kyushu-u.ac.jp}
}

\begin{abstract}
By employing the gradient/Wilson flow, we derive a universal formula that
expresses a correctly normalized flavor non-singlet axial-vector current of
quarks. The formula is universal in the sense that it holds independently of
regularization and especially holds with lattice regularization. It is also
confirmed that, in the lowest non-trivial order of perturbation theory, the
triangle diagram containing the formula and two flavor non-singlet vector
currents possesses non-local structure that is compatible with the triangle
anomaly.
\end{abstract}
\subjectindex{B01, B31, B32, B38}
\maketitle

\section{Introduction}
\label{sec:1}
In this paper, we consider a well-studied problem, how to construct a correctly
(or canonically) normalized flavor non-singlet axial-vector current of
quarks,\footnote{See Ref.~\cite{Aoki:2013ldr} and references cited therein
for various methods.} in a new light using the gradient/Wilson
flow~\cite{Luscher:2010iy,Luscher:2011bx,Luscher:2013cpa}.\footnote{A strategy
to determine a non-perturbative renormalization constant of the axial-vector
current on the basis of the axial Ward--Takahashi relation in the flowed
system has been developed in~Ref.~\cite{Luscher:2013cpa}. See also
Ref.~\cite{Shindler:2013bia} for a detailed study of the axial Ward--Takahashi
relation in the flowed system. In these papers, the flowed fields are
employed as a ``probe'' rather than to construct the axial-vector current
itself. In the present paper, we instead construct the axial-vector current
from the flowed fields through the small flow-time expansion.} An axial-vector
current $j_{5\mu}^A(x)$ is said to be correctly normalized, if it fulfills the
Ward--Takahashi relation associated with the flavor chiral symmetry. That
partially conserved axial current (PCAC) relation is
\begin{align}
   &\left\langle
   \partial_\mu j_{5\mu}^A(x)
   \psi(y)\Bar{\psi}(z)\right\rangle
   -\left\langle\Bar{\psi}(x)\gamma_5\{t^A,M_0\}\psi(x)
   \psi(y)\Bar{\psi}(z)\right\rangle
\notag\\
   &=-\delta^4(y-x)\gamma_5t^A\left\langle\psi(y)\Bar{\psi}(z)\right\rangle
   -\delta^4(z-x)\left\langle\psi(y)\Bar{\psi}(z)\right\rangle\gamma_5 t^A,
\label{eq:(1.1)}
\end{align}
where $\psi$ and~$\Bar{\psi}$ are quark and anti-quark fields and $t^A$ ($A=1$,
\dots, $N_f^2-1$) denotes the generator of the flavor $SU(N_f)$ group. $M_0$ is
the (flavor-diagonal) bare quark mass matrix. This relation says that the
axial-vector current~$j_{5\mu}^A(x)$ generates the axial part of the flavor
symmetry~$SU(N_f)_L\times SU(N_f)_R$ in the correct magnitude. The construction
of such a composite operator is, however, not straightforward because a
regulator that manifestly preserves the chiral symmetry does not easily come to
hand. The only known explicit examples of such chiral-symmetry-preserving
regularization are the domain-wall lattice
fermion~\cite{Kaplan:1992bt,Shamir:1993zy,Furman:1994ky} and the overlap
lattice fermion~\cite{Neuberger:1997fp,Neuberger:1998wv}, both satisfy the
Ginsparg--Wilson relation~\cite{Ginsparg:1981bj,Hasenfratz:1998jp,%
Hasenfratz:1998ri,Luscher:1998pqa}.

For example, if one uses dimensional regularization with complex
dimension~$D=4-2\epsilon$ and
\begin{equation}
   \gamma_5=\gamma_0\gamma_1\gamma_2\gamma_3,
\label{eq:(1.2)}
\end{equation}
a simple expression for the axial-vector current
\begin{equation}
   \Bar{\psi}(x)\gamma_\mu\gamma_5 t^A\psi(x)
\label{eq:(1.3)}
\end{equation}
does not fulfill Eq.~\eqref{eq:(1.1)}. Instead one finds (see
Ref.~\cite{Collins:1984xc}, for example)\footnote{Here, $g_0$ is the bare gauge
coupling. This expression is for a fermion in a generic gauge
representation~$R$ of a gauge group~$G$. The quadratic Casimir~$C_2(R)$ is
defined by~$T^aT^a=-C_2(R)\mathbb{1}$ from anti-Hermitian generators~$T^a$
of~$G$; we normalize generators as~$\tr_R(T^aT^b)=-T(R)\delta^{ab}$. For the
fundamental representation~$N$ of~$G=SU(N)$, the conventional normalization is
$T(N)=1/2$ and~$C_2(N)=(N^2-1)/(2N)$; $C_2(R)=4/3$ for quarks.}
\begin{align}
   &\left\langle
   \partial_\mu\left[\Bar{\psi}(x)\gamma_\mu\gamma_5t^A\psi(x)\right]
   \psi(y)\Bar{\psi}(z)\right\rangle
   -\left\langle\Bar{\psi}(x)\gamma_5\{t^A,M_0\}\psi(x)
   \psi(y)\Bar{\psi}(z)\right\rangle
\notag\\
   &=-\delta^4(y-x)\gamma_5t^A\left\langle\psi(y)\Bar{\psi}(z)\right\rangle
   -\delta^4(z-x)\left\langle\psi(y)\Bar{\psi}(z)\right\rangle\gamma_5t^A
\notag\\
   &\qquad{}
   +\frac{g_0^2}{(4\pi)^2}C_2(R)4
   \left\langle
   \partial_\mu\left[\Bar{\psi}(x)\gamma_\mu\gamma_5t^A\psi(x)\right]
   \psi(y)\Bar{\psi}(z)\right\rangle
\notag\\
   &\qquad\qquad{}
   -\frac{g_0^2}{(4\pi)^2}C_2(R)8
   \left\langle
   \Bar{\psi}(x)\gamma_5\{t^A,M_0\}\psi(x)
   \psi(y)\Bar{\psi}(z)\right\rangle+O(g_0^4)
\label{eq:(1.4)}
\end{align}
for $\epsilon\to0$. This relation shows that under dimensional regularization,
the correctly normalized axial-vector current is not Eq.~\eqref{eq:(1.3)} but
\begin{equation}
   j_{5\mu}^A(x)
   =\left[1-\frac{g_0^2}{(4\pi)^2}C_2(R)4+O(g_0^4)\right]
   \Bar{\psi}(x)\gamma_\mu\gamma_5t^A\psi(x)
\label{eq:(1.5)}
\end{equation}
in conjunction with a redefinition of the pseudo-scalar density,
\begin{equation}
   \left[1-\frac{g_0^2}{(4\pi)^2}C_2(R)8+O(g_0^4)\right]
   \Bar{\psi}(x)\gamma_5\{t^A,M_0\}\psi(x).
\label{eq:(1.6)}
\end{equation}
The relation between the correctly normalized axial-vector current and a bare
axial-vector current is regularization-dependent and generally receives
radiative corrections in all orders of perturbation theory. If one changes
regularization (to lattice regularization with a particular discretization of
the Dirac operator, for example), one has to compute the corresponding relation
anew.

In the present paper, instead, we derive a single ``universal formula'' that is
supposed to hold for \emph{any\/} regularization. Our result is
\begin{equation}
   j_{5\mu}^A(x)
   =\lim_{t\to0}\left\{1+
   \frac{\Bar{g}(1/\sqrt{8t})^2}{(4\pi)^2}C_2(R)
   \left[-\frac{1}{2}+\ln(432)\right]\right\}
   \mathring{\Bar{\chi}}(t,x)\gamma_\mu\gamma_5t^A\mathring{\chi}(t,x).
\label{eq:(1.7)}
\end{equation}
Here, $\Bar{g}(q)$ is the running gauge coupling in the minimal subtraction
(MS) scheme at the renormalization scale~$q$;\footnote{For completeness, we
quote the related formulas: The running gauge coupling is defined by
\begin{equation}
   q\frac{\mathrm{d}\Bar{g}(q)}{\mathrm{d}q}=\beta(\Bar{g}(q)),\qquad
   \Bar{g}(q=\mu)=g,
\label{eq:(1.8)}
\end{equation}
where the beta function is given from the renormalization constant~$Z$ in the
MS scheme in
\begin{equation}
   g_0^2=\mu^{2\epsilon}g^2Z,\qquad
   Z=1-\frac{1}{\epsilon}z^{(1)}-\frac{1}{\epsilon^2}z^{(2)}+\dotsb,
\label{eq:(1.9)}
\end{equation}
by
\begin{equation}
   \frac{\beta}{g}=-\epsilon-g^2\frac{d}{dg^2}z^{(1)}.
\label{eq:(1.10)}
\end{equation}
}
$\mathring{\chi}(t,x)$ and~$\mathring{\Bar{\chi}}(t,x)$ are quark and
anti-quark fields evolved by flow equations which will be elucidated below.
Because of theorems proven in~Refs.~\cite{Luscher:2011bx,Luscher:2013cpa}, the
composite operator in the right-hand side of~Eq.~\eqref{eq:(1.7)} is a
renormalized quantity although it is constructed from bare quark fields in a
well-defined manner. As far as one carries out the parameter renormalization
properly, such a renormalized quantity must be independent of the chosen
regularization. In this sense, the formula is ``universal.'' The formula is
also usable with lattice regularization and could be applied for the
computation of, say, the pion decay constant.

Since $\Bar{g}(1/\sqrt{8t})\to0$ for~$t\to0$ because of the asymptotic freedom,
the formula~\eqref{eq:(1.7)} can further be simplified
to~$j_{5\mu}^A(x)=\lim_{t\to0}%
\mathring{\Bar{\chi}}(t,x)\gamma_\mu\gamma_5t^A\mathring{\chi}(t,x)$. Although
this is mathematically correct, practically one cannot simply take the
$t\to0$~limit in lattice Monte Carlo simulations for example.
In~Eq.~\eqref{eq:(1.7)}, it is supposed that the regulator is removed (after
making the parameter renormalization) and the lower end of the physical flow
time~$t$ is limited by the lattice spacing as~$a\ll\sqrt{8t}$. Thus the
asymptotic $t\to0$ behavior in~Eq.~\eqref{eq:(1.7)} will be useful to find the
extrapolation for~$t\to0$.

The above universal formula for the axial-vector current is quite analogous to
universal formulas for the energy--momentum tensor---the Noether current
associated with the translational
invariance---on the basis of the gradient/Wilson
flow~\cite{Suzuki:2013gza,Makino:2014taa,Makino:2014sta,Suzuki:2015fka};
see also Ref.~\cite{DelDebbio:2013zaa}. The motivation is also similar: Since
lattice regularization breaks the translational invariance, the construction
of the energy--momentum tensor with lattice regularization is not
straightforward~\cite{Caracciolo:1988hc,Caracciolo:1989pt}. The universal
formulas
in~Refs.~\cite{Suzuki:2013gza,Makino:2014taa,Makino:2014sta,Suzuki:2015fka},
because they are thought to be regularization independent, may be used with
lattice regularization. The validity of the formulas has been tested by
employing Monte Carlo simulations~\cite{Asakawa:2013laa,Kitazawa:2014uxa} and
the $1/N$ expansion~\cite{Makino:2014cxa} (see also, Ref.~\cite{Aoki:2014dxa}
for the $1/N$ expansion of the gradient flow).

A main technical difference in the derivation of~Eq.~\eqref{eq:(1.7)} from the
derivation of the universal formulas for the energy--momentum tensor
in~Refs.~\cite{Suzuki:2013gza,Makino:2014taa,Makino:2014sta,Suzuki:2015fka} is
that only the knowledge of the one-loop expression in dimensional
regularization~\eqref{eq:(1.5)} will be used in what follows. On the other
hand,
in~Refs.~\cite{Suzuki:2013gza,Makino:2014taa,Makino:2014sta,Suzuki:2015fka},
full-order perturbative expressions of the correctly normalized
energy--momentum tensor (which is readily obtained by dimensional
regularization) were used. Although it must be possible to arrive
at~Eq.~\eqref{eq:(1.7)} starting from an ``ideal'' axial-vector current
obtained from the chiral-symmetry-preserving lattice fermions by the Noether
method, a much simpler one-loop expression such as~Eq.~\eqref{eq:(1.5)} is
sufficient. This point, we think is technically very interesting.\footnote{The
idea that the knowledge of a one-loop expression such as~Eq.~\eqref{eq:(1.5)}
would be sufficient to find the universal formula has emerged through
discussion with David B. Kaplan. We would like to thank him.}

\section{Gradient/Wilson flow}
\label{sec:2}
The gradient/Wilson flow~\cite{Luscher:2010iy,Luscher:2011bx,Luscher:2013cpa}
is an evolution of field configurations according to flow equations with a
fictitious time~$t$. For the gauge field~$A_\mu(x)$, the flow is defined
by~\cite{Luscher:2010iy,Luscher:2011bx}
\begin{equation}
   \partial_tB_\mu(t,x)=D_\nu G_{\nu\mu}(t,x)
   +\alpha_0D_\mu\partial_\nu B_\nu(t,x),\qquad
   B_\mu(t=0,x)=A_\mu(x),
\label{eq:(2.1)}
\end{equation}
where $G_{\mu\nu}(t,x)$ is the field strength of the flowed gauge
field~$B_\mu(t,x)$,
\begin{equation}
   G_{\mu\nu}(t,x)
   =\partial_\mu B_\nu(t,x)-\partial_\nu B_\mu(t,x)
   +[B_\mu(t,x),B_\nu(t,x)],
\label{eq:(2.2)}
\end{equation}
and the covariant derivative on the gauge field is
\begin{equation}
   D_\mu=\partial_\mu+[B_\mu,\cdot].
\label{eq:(2.3)}
\end{equation}
The second term in the right-hand side of~Eq.~\eqref{eq:(2.1)} is a
``gauge-fixing term'' that is introduced to provide the Gaussian damping
factor (see below) to the gauge mode. Although it breaks the gauge covariance,
it can be shown that any gauge-invariant quantity is independent of the
``gauge parameter''~$\alpha_0$~\cite{Luscher:2010iy}.

For the quark fields, $\psi(x)$ and~$\Bar{\psi}(x)$, the flow is defined
by~\cite{Luscher:2013cpa}
\begin{align}
   &\partial_t\chi(t,x)=\left[\Delta-\alpha_0\partial_\mu B_\mu(t,x)\right]
   \chi(t,x),\qquad
   \chi(t=0,x)=\psi(x),
\label{eq:(2.4)}
\\
   &\partial_t\Bar{\chi}(t,x)
   =\Bar{\chi}(t,x)
   \left[\overleftarrow{\Delta}
   +\alpha_0\partial_\mu B_\mu(t,x)\right],
   \qquad\Bar{\chi}(t=0,x)=\Bar{\psi}(x),
\label{eq:(2.5)}
\end{align}
where covariant derivatives for the flowed quark field, $\chi(t,x)$
and~$\Bar{\chi}(t,x)$, are
\begin{align}
   &\Delta=D_\mu D_\mu,\qquad D_\mu=\partial_\mu+B_\mu,
\label{eq:(2.6)}
\\
   &\overleftarrow{\Delta}=\overleftarrow{D}_\mu\overleftarrow{D}_\mu,
   \qquad\overleftarrow{D}_\mu=\overleftarrow{\partial}_\mu-B_\mu.
\label{eq:(2.7)}
\end{align}
For the implementation of these flow equations in lattice gauge theory, see
Refs.~\cite{Luscher:2010iy,Luscher:2013cpa}.

The initial values in above flow equations, $A_\mu(x)$, $\psi(x)$,
and~$\Bar{\psi}(x)$, are quantum fields being subject to the functional
integral. One can develop~\cite{Luscher:2010iy,Luscher:2011bx,Luscher:2013cpa}
perturbation theory for quantum correlation functions of the flowed fields,
$B_\mu(t,x)$, $\chi(t,x)$, and~$\Bar{\chi}(t,x)$. For example, the tree-level
propagator of the flowed gauge field (in the ``Feynman gauge'' in which
$\lambda_0=\alpha_0=1$, where $\lambda_0$ is the conventional gauge-fixing
parameter) is\footnote{Throughout this paper, we use the abbreviation
\begin{equation}
   \int_p\equiv\int\frac{\mathrm{d}^Dp}{(2\pi)^D}.
\label{eq:(2.8)}
\end{equation}
}
\begin{equation}
   \left\langle B_\mu^a(t,x)B_\nu^b(s,y)\right\rangle
   =g_0^2\delta^{ab}\delta_{\mu\nu}\int_p\mathrm{e}^{ip(x-y)}
   \frac{\mathrm{e}^{-(t+s)p^2}}{p^2}.
\label{eq:(2.9)}
\end{equation}
Similarly, the tree-level quark propagator is
\begin{equation}
   \left\langle\chi(t,x)\Bar{\chi}(s,y)\right\rangle
   =\int_p\mathrm{e}^{ip(x-y)}\frac{\mathrm{e}^{-(t+s)p^2}}{i\Slash{p}+M_0}.
\label{eq:(2.10)}
\end{equation}
The details of the perturbation theory (the flow Feynman rule) are summarized
in~Ref.~\cite{Makino:2014taa}.

A remarkable feature of the gradient/Wilson flow is its ultraviolet (UV)
finiteness~\cite{Luscher:2011bx,Luscher:2013cpa}. Correlation functions of
the flowed gauge field become UV finite \emph{without\/} the wave function
renormalization (if one makes the parameter renormalization). This finiteness
persists even for local products, i.e., composite operators. The flowed quark
field, on the other hand requires the wave function renormalization. However,
once the elementary flowed quark field is multiplicatively renormalized, any
composite operator of it becomes UV finite. Thus, if one introduces the
combinations~\cite{Makino:2014taa}\footnote{Here $\dim(R)$ is the dimension
of the representation~$R$ of the gauge group to which the fermion belongs
($\dim(R)=3$ for quarks). The coefficients in these expressions are chosen
so that $\mathring{\chi}(t,x)\to\chi(t,x)$
and~$\mathring{\Bar{\chi}}(t,x)\to\Bar{\chi}(t,x)$
for~$t\to0$~\cite{Makino:2014taa}.}
\begin{align}
   \mathring{\chi}(t,x)
   &=\sqrt{\frac{-2\dim(R)N_f}
   {(4\pi)^2t^2
   \left\langle\Bar{\chi}(t,x)\overleftrightarrow{\Slash{D}}\chi(t,x)
   \right\rangle}}
   \,\chi(t,x),
\label{eq:(2.11)}
\\
   \mathring{\Bar{\chi}}(t,x)
   &=\sqrt{\frac{-2\dim(R)N_f}
   {(4\pi)^2t^2
   \left\langle\Bar{\chi}(t,x)\overleftrightarrow{\Slash{D}}\chi(t,x)
   \right\rangle}}
   \,\Bar{\chi}(t,x),
\label{eq:(2.12)}
\end{align}
where
\begin{equation}
   \overleftrightarrow{D}_\mu\equiv D_\mu-\overleftarrow{D}_\mu,
\label{eq:(2.13)}
\end{equation}
then wave function renormalization factors are canceled out in
$\mathring{\chi}(t,x)$ and~$\mathring{\Bar{\chi}}(t,x)$ and any local product
of them becomes a UV-finite renormalized operator. Thus, if one can express a
composite operator in terms of local products of flowed fields, $B_\mu(t,x)$,
$\mathring{\chi}(t,x)$, and~$\mathring{\Bar{\chi}}(t,x)$, then the expression
provides a universal formula for the composite operator. What we shall do is to
apply this idea to the composite operator in~Eq.~\eqref{eq:(1.5)}. An explicit
method to rewrite a composite operator in terms of local products of the flowed
fields is given by the short flow-time expansion~\cite{Luscher:2011bx} that is
the subject of the next section.

\section{Small flow-time expansion}
\label{sec:3}
We take a would-be axial-vector current composed of the flowed quark field
\begin{equation}
   \Bar{\chi}(t,x)\gamma_\mu\gamma_5t^A\chi(t,x),
\label{eq:(3.1)}
\end{equation}
and consider its $t\to0$ limit. As discussed in~Ref.~\cite{Luscher:2011bx}, a
local product of flowed fields in the $t\to0$~limit can be expressed by an
asymptotic series of \emph{local\/} composite operators of fields at zero
flow time with increasing mass dimensions. In the present case, because of
symmetry, we have
\begin{equation}
   \Bar{\chi}(t,x)\gamma_\mu\gamma_5t^A\chi(t,x)
   =c(t)\Bar{\psi}(x)\gamma_\mu\gamma_5t^A\psi(x)+O(t)
\label{eq:(3.2)}
\end{equation}
for~$t\to0$.

\begin{figure}[ht]
\begin{center}
\includegraphics[width=0.11\columnwidth]{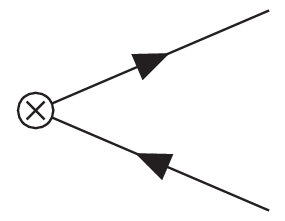}
\caption{Diagram a}
\label{fig:1}
\end{center}
\end{figure}

\begin{figure}[ht]
\begin{minipage}{0.3\columnwidth}
\begin{center}
\includegraphics[width=0.7\columnwidth]{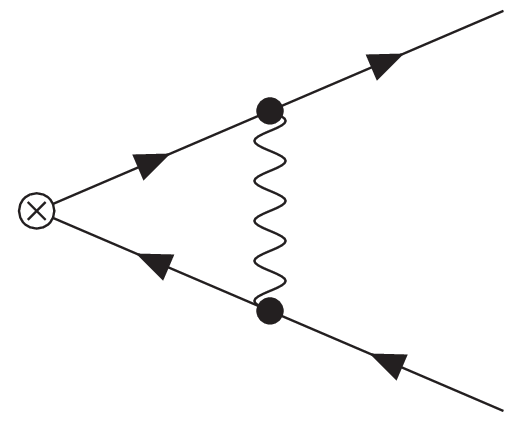}
\caption{Diagram d}
\label{fig:2}
\end{center}
\end{minipage}
\begin{minipage}{0.3\columnwidth}
\begin{center}
\includegraphics[width=0.7\columnwidth]{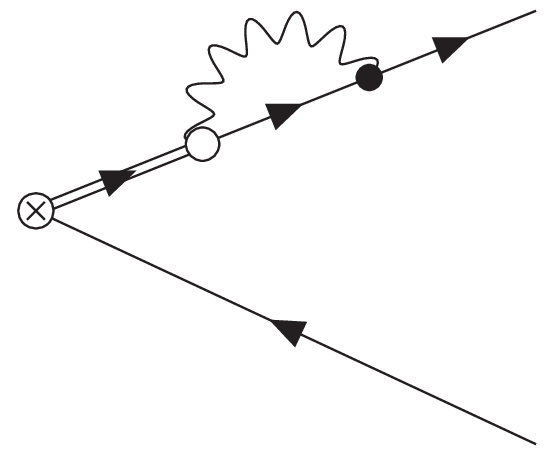}
\caption{Diagram f}
\label{fig:3}
\end{center}
\end{minipage}
\begin{minipage}{0.3\columnwidth}
\begin{center}
\includegraphics[width=0.7\columnwidth]{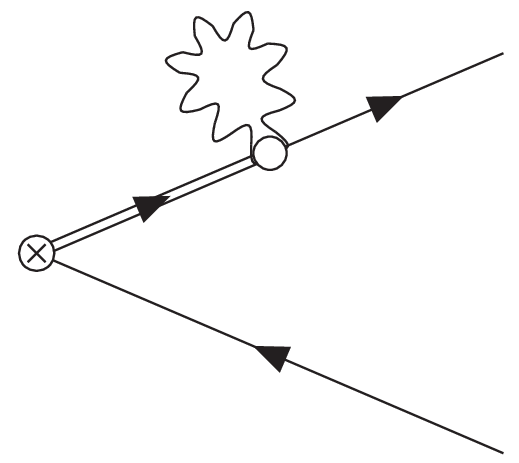}
\caption{Diagram g}
\label{fig:4}
\end{center}
\end{minipage}
\end{figure}

\begin{figure}[ht]
\begin{minipage}{0.3\columnwidth}
\begin{center}
\includegraphics[width=0.7\columnwidth]{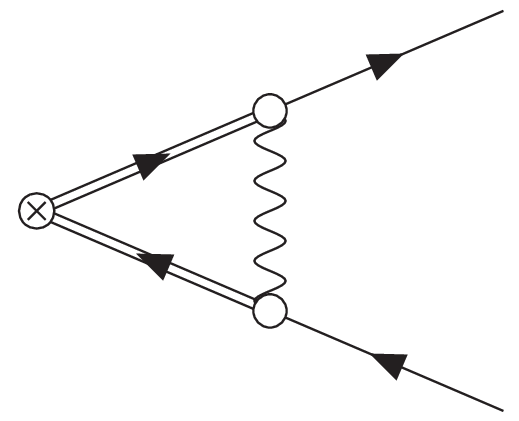}
\caption{Diagram b}
\label{fig:5}
\end{center}
\end{minipage}
\begin{minipage}{0.3\columnwidth}
\begin{center}
\includegraphics[width=0.7\columnwidth]{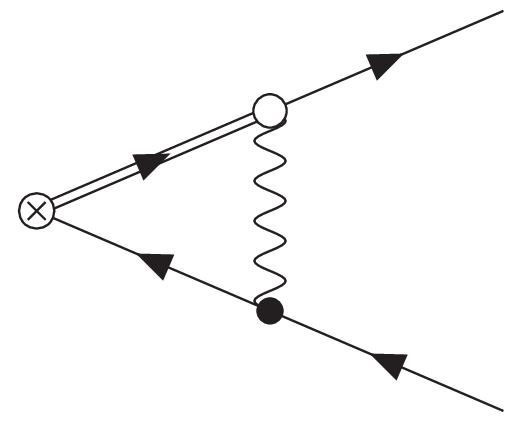}
\caption{Diagram c}
\label{fig:6}
\end{center}
\end{minipage}
\begin{minipage}{0.3\columnwidth}
\begin{center}
\includegraphics[width=0.7\columnwidth]{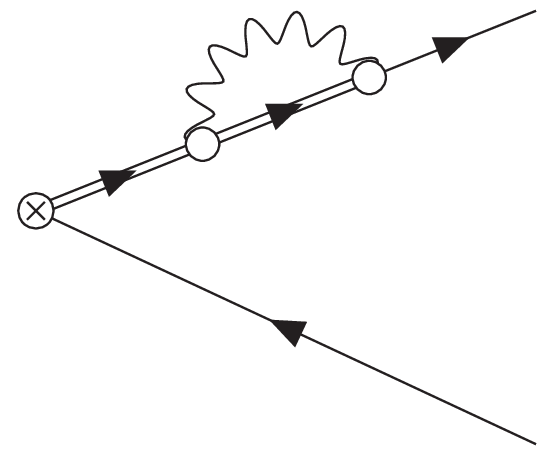}
\caption{Diagram e}
\label{fig:7}
\end{center}
\end{minipage}
\end{figure}

The expansion can be worked out in perturbation theory; the
coefficient~$c(t)$ can be found by computing the correlation function
\begin{equation}
   \left\langle\Bar{\chi}(t,x)\gamma_\mu\gamma_5t^A\chi(t,x)
   \psi(y)\Bar{\psi}(z)\right\rangle
\label{eq:(3.3)}
\end{equation}
and comparing its $t\to0$ behavior with
\begin{equation}
   \left\langle\Bar{\psi}(x)\gamma_\mu\gamma_5t^A\psi(x)
   \psi(y)\Bar{\psi}(z)\right\rangle.
\label{eq:(3.4)}
\end{equation}
To one-loop order, there are seven 1PI ``flow Feynman diagrams'' which
contribute to~Eq.~\eqref{eq:(3.3)}; they are depicted
in~Figs~\ref{fig:1}--\ref{fig:7} according to the convention
in~Ref.~\cite{Makino:2014taa}.
Among these diagrams, diagrams b, c, and~e
in~Figs.~\ref{fig:5}--\ref{fig:7} are irrelevant for the coefficient~$c(t)$
in~Eq.~\eqref{eq:(3.2)} because they are proportional to the momentum of the
external quark line. Then, writing Eq.~\eqref{eq:(3.3)} as
\begin{equation}
   \int\frac{\mathrm{d}^Dp}{(2\pi)^D}\int\frac{\mathrm{d}^Dq}{(2\pi)^D}\,
   \mathrm{e}^{ip(y-x)}\mathrm{e}^{iq(x-z)}\frac{1}{i\Slash{p}+M_0}
   \mathcal{M}_{5\mu}^A(p,q;t)
   \frac{1}{i\Slash{q}+M_0},
\label{eq:(3.5)}
\end{equation}
the contribution of diagram~a is
$\mathcal{M}_{5\mu}^A(0,0;t)=\gamma_\mu\gamma_5t^A$. For one-loop diagrams, we
work with dimensional regularization [with the prescription~\eqref{eq:(1.2)}].
The contribution of each diagram to~$\mathcal{M}_{5\mu}^A(0,0;t)$ is tabulated
in~Table~\ref{table:1}.
\begin{table}[ht]
\caption{Contribution of each diagram to~$\mathcal{M}_{5\mu}^A(0,0;t)$
in~Eq.~\eqref{eq:(3.5)} in units
of~$\frac{g_0^2}{(4\pi)^2}C_2(R)\gamma_\mu\gamma_5t^A$.}
\label{table:1}
\begin{center}
\renewcommand{\arraystretch}{2.2}
\setlength{\tabcolsep}{20pt}
\begin{tabular}{cr}
\toprule
d  & $(-1)\left[\dfrac{1}{\epsilon}+\ln(8\pi t)+\dfrac{7}{2}\right]$ \\
f  & $2\left[\dfrac{1}{\epsilon}+\ln(8\pi t)+1\right]$ \\
g  & $(-4)\left[\dfrac{1}{\epsilon}+\ln(8\pi t)+\dfrac{1}{2}\right]$ \\
\bottomrule
\end{tabular}
\end{center}
\end{table}
Combining all the contributions, we have
\begin{equation}
    \mathcal{M}_{5\mu}^A(0,0;t)
   =\left\{1+\frac{g_0^2}{(4\pi)^2}C_2(R)(-3)
   \left[\frac{1}{\epsilon}+\ln(8\pi t)+\frac{7}{6}\right]\right\}
   \gamma_\mu\gamma_5t^A.
\label{eq:(3.6)}
\end{equation}
Invoking a facile method explaining in~Ref.~\cite{Makino:2014taaa}, this
implies the small flow-time expansion is given by
\begin{align}
   &\Bar{\chi}(t,x)\gamma_\mu\gamma_5t^A\chi(t,x)
\notag\\
   &=\left\{1+
   \frac{g_0^2}{(4\pi)^2}C_2(R)(-3)
   \left[\frac{1}{\epsilon}+\ln(8\pi t)+\frac{7}{6}\right]\right\}
   \Bar{\psi}(x)\gamma_\mu\gamma_5t^A\psi(x)
   +O(t).
\label{eq:(3.7)}
\end{align}

Next we express this in terms of the ``ringed'' fields
in~Eqs.~\eqref{eq:(2.11)} and~\eqref{eq:(2.12)}. Using the result
of~Ref.~\cite{Makino:2014taa} that
\begin{equation}
   \frac{-2\dim(R)N_{\mathrm{f}}}
   {(4\pi)^2t^2\left\langle\Bar{\chi}(t,x)
   \overleftrightarrow{\Slash{D}}\chi(t,x)\right\rangle}
   =Z(\epsilon)
   \left\{1+\frac{g^2}{(4\pi)^2}C_2(R)\left[3\frac{1}{\epsilon}-\Phi(t)\right]
   +O(g^4)+O(t)\right\},
\label{eq:(3.8)}
\end{equation}
where
\begin{equation}
   Z(\epsilon)\equiv\frac{1}{(8\pi t)^\epsilon},\qquad
   \Phi(t)\equiv-3\ln(8\pi\mu^2t)+\ln(432),
\label{eq:(3.9)}
\end{equation}
we have for~$\epsilon\to0$,
\begin{align}
   &\mathring{\Bar{\chi}}(t,x)\gamma_\mu\gamma_5t^A\mathring{\chi}(t,x)
\notag\\
   &=\left\{1+
   \frac{g^2}{(4\pi)^2}C_2(R)
   \left[-\frac{7}{2}-\ln(432)\right]+O(g^4)\right\}
   \Bar{\psi}(x)\gamma_\mu\gamma_5t^A\psi(x)
   +O(t),
\label{eq:(3.10)}
\end{align}
where we have set~$g_0^2=\mu^{2\epsilon}g^2$.

Inverting this relation for~$t\to0$ and plugging it into~Eq.~\eqref{eq:(1.6)},
we have
\begin{equation}
   j_{5\mu}^A(x)
   =\left\{1+
   \frac{g^2}{(4\pi)^2}C_2(R)
   \left[-\frac{1}{2}+\ln(432)\right]+O(g^4)\right\}
   \mathring{\Bar{\chi}}(t,x)\gamma_\mu\gamma_5t^A\mathring{\chi}(t,x)
   +O(t).
\label{eq:(3.11)}
\end{equation}
As the Ward--Takahashi relation~\eqref{eq:(1.1)} shows, $j_{5\mu}^A(x)$ in the
left-hand side of this relation must be UV finite. The right-hand side of
this relation is certainly UV finite, as all $1/\epsilon$ singularities are
cancelled out.

Finally, we invoke a renormalization group argument. By applying the operation
\begin{equation}
   \left(\mu\frac{\partial}{\partial\mu}\right)_0
\label{eq:(3.12)}
\end{equation}
to the both sides of~Eq.~\eqref{eq:(1.1)}, where the subscript~$0$ implies the
bare quantities are kept fixed, we have
$(\mu\partial/\partial\mu)_0j_{5\nu}^A(x)=0$.\footnote{Since
$j_{5\nu}^A(x)$ is the unique gauge invariant flavor non-singlet dimension~$3$
axial-vector operator,
$(\mu\partial/\partial\mu)_0\partial_\nu j_{5\nu}^A(x)=0$ implies this.}
We also have $(\mu\partial/\partial\mu)_0
\mathring{\Bar{\chi}}(t,x)\gamma_\mu\gamma_5t^A\mathring{\chi}(t,x)=0$, because
the flowed quark fields are certain (although very complicated) combinations of
bare fields. Therefore, the quantity in the curly brackets
in~Eq.~\eqref{eq:(3.10)} is independent of the renormalization scale~$\mu$.
This implies that if one uses the running gauge coupling~$\Bar{g}(q)$ in place
of~$g$, the expression is independent of the scale~$q$. Thus, we set
$q=1/\sqrt{8t}$ as a particular choice. Since $\Bar{g}(1/\sqrt{8t})\to0$
for~$t\to0$, the perturbative computation is justified in the $t\to0$ limit
which also eliminates the $O(t)$~term in~Eq.~\eqref{eq:(3.11)}. In this way, we
arrive at the universal formula~\eqref{eq:(1.7)}.

An argument similar to the above may be repeated for the flavor non-singlet
pseudo-scalar density that fulfills the PCAC relation,
\begin{align}
   &\left\langle
   \partial_\mu j_{5\mu}^A(x)
   \psi(y)\Bar{\psi}(z)\right\rangle
   -\left\langle
   \left\{\Bar{\psi}\gamma_5\{t^A,M\}\psi\right\}_R(x)
   \psi(y)\Bar{\psi}(z)\right\rangle
\notag\\
   &=-\delta^4(y-x)\gamma_5t^A\left\langle\psi(y)\Bar{\psi}(z)\right\rangle
   -\delta^4(z-x)\left\langle\psi(y)\Bar{\psi}(z)\right\rangle\gamma_5 t^A.
\label{eq:(3.13)}
\end{align}
Omitting all details of the calculation, the final result is given by
\begin{align}
   &\left\{\Bar{\psi}\gamma_5\{t^A,M\}\psi\right\}_R(x)
\notag\\
   &=\lim_{t\to0}\left\{1+
   \frac{\Bar{g}(1/\sqrt{8t})^2}{(4\pi)^2}C_2(R)
   \left[3\ln\pi+2
   +\ln(432)\right]\right\}
   \mathring{\Bar{\chi}}(t,x)\gamma_5\{t^A,\Bar{M}(1/\sqrt{8t})\}
   \mathring{\chi}(t,x).
\label{eq:(3.14)}
\end{align}
In these expressions, the MS or $\overline{\text{MS}}$ scheme is assumed both
for the renormalized mass matrix~$M$ and the renormalized pseudo-scalar
density; $\Bar{M}(q)$ is the running mass matrix at the renormalization
scale~$q$.

\section{Triangle anomaly}
\label{sec:4}
The matrix element of the above flavor non-singlet axial-vector current is
relevant for the pseudo-scalar meson decay process by weak interaction. For the
process $\pi^0\to2\gamma$, the triangle diagram containing one axial-vector
current and two vector currents is important~\cite{Adler:1969gk,Bell:1969ts}.
It is thus of great interest to examine the three-point
function\footnote{Here, we supposed that $[t^A,t^B]=[t^A,t^C]=0$.}
\begin{equation}
   \mathcal{T}_{\mu\nu\rho}^{ABC}(p,q)
   \equiv\int\mathrm{d}^Dx\int\mathrm{d}^Dy\,
   \mathrm{e}^{ipx}\,\mathrm{e}^{iqy}\,\frac{1}{4}\left[
   \left\langle j_{5\mu}^A(0)j_\nu^B(x)j_\rho^C(y)\right\rangle
   +(B\leftrightarrow C)\right],
\label{eq:(4.1)}
\end{equation}
where $j_{5\mu}^A(0)$ is defined by our universal formula~\eqref{eq:(1.7)} and
$j_\nu^B(x)$ and~$j_\rho^C(y)$ are flavor non-singlet vector currents
\begin{equation}
   j_\nu^B(y)\equiv\Bar{\psi}(y)\gamma_\nu t^B\psi(y),\qquad
   j_\rho^C(z)\equiv\Bar{\psi}(z)\gamma_\rho t^C\psi(z).
\label{eq:(4.2)}
\end{equation}
Usually, conventional regularization preserves the flavor vector symmetry and a
naive expression of the vector current is correctly normalized.

In the present paper, we investigate whether~\eqref{eq:(4.1)} reproduces the
correct triangle (or axial) anomaly in the lowest non-trivial order of
perturbation theory. We work with massless theory for simplicity. In the lowest
order of perturbation theory, using Eqs.~\eqref{eq:(1.7)}, \eqref{eq:(2.11)},
\eqref{eq:(2.12)}, \eqref{eq:(3.8)}, and finally \eqref{eq:(2.10)},
Eq.~\eqref{eq:(4.1)} before taking the $t\to0$ limit is given
by\footnote{$\tr_R(1)$ denotes the trace of the unit matrix over the gauge
representation index; $\tr_R(1)=3$ for quarks.}
\begin{align}
   \left.\mathcal{T}_{\mu\nu\rho}^{ABC}(p,q)\right|_{O(g^0)}
   &=-\frac{i}{4}\tr_R(1)\tr\left(t^A\{t^B,t^C\}\right)
\notag\\
   &\qquad{}
   \times\int_\ell\tr\biggl[
   \gamma_\mu\gamma_5\frac{1}{\Slash{\ell}+\Slash{p}}
   \gamma_\nu\frac{1}{\Slash{\ell}}\gamma_\rho
   \frac{1}{\Slash{\ell}-\Slash{q}}
   \mathrm{e}^{-t(\ell+p)^2}\mathrm{e}^{-t(\ell-q)^2}
\notag\\
   &\qquad\qquad\qquad{}
   +\gamma_\mu\gamma_5\frac{1}{\Slash{\ell}+\Slash{q}}
   \gamma_\rho\frac{1}{\Slash{\ell}}\gamma_\nu
   \frac{1}{\Slash{\ell}-\Slash{p}}
   \mathrm{e}^{-t(\ell+q)^2}\mathrm{e}^{-t(\ell-p)^2}\biggr].
\label{eq:(4.3)}
\end{align}
The total divergence of the axial-vector current, i.e., the triangle anomaly,
is thus given by
\begin{align}
   &i(p+q)_\mu\left.\mathcal{T}_{\mu\nu\rho}^{ABC}(p,q)\right|_{O(g^0)}
\notag\\
   &=-\frac{1}{4}\tr_R(1)\tr\left(t^A\{t^B,t^C\}\right)
\notag\\
   &\qquad{}\times\int_\ell\tr\gamma_5\left[
   \frac{1}{\Slash{\ell}+\Slash{p}}
   \gamma_\nu\frac{1}{\Slash{\ell}}\gamma_\rho
   \mathrm{e}^{-t(\ell+p)^2}\mathrm{e}^{-t(\ell-q)^2}
   -\frac{1}{\Slash{\ell}}\gamma_\nu
   \frac{1}{\Slash{\ell}-\Slash{p}}\gamma_\rho
   \mathrm{e}^{-t(\ell-p)^2}\mathrm{e}^{-t(\ell+q)^2}
   \right]
\notag\\
   &\qquad\qquad{}
   +(p\leftrightarrow q,\nu\leftrightarrow\rho).
\label{eq:(4.4)}
\end{align}
It is interesting to note that if there were no Gaussian damping factors which
result from the flow of the quark fields in this expression, a naive shift of
the loop momentum makes this expression vanish as the well-known
case~\cite{Adler:1969gk,Bell:1969ts}. The reality is that there are Gaussian
damping factors and, in the $t\to0$ limit, we have the following non-zero
result:
\begin{align}
   i(p+q)_\mu\left.\mathcal{T}_{\mu\nu\rho}^{ABC}(p,q)\right|_{O(g^0)}
   &=-2t\int_\ell\frac{\mathrm{e}^{-2t\ell^2}}{\ell^2}
   \tr_R(1)\tr\left(t^A\{t^B,t^C\}\right)
   \epsilon_{\alpha\nu\beta\rho}p_\alpha q_\beta
\notag\\
   &=-\frac{1}{16\pi^2}\tr_R(1)\tr\left(t^A\{t^B,t^C\}\right)
   \epsilon_{\alpha\nu\beta\rho}p_\alpha q_\beta.
\label{eq:(4.5)}
\end{align}
However, this is \emph{not\/} quite identical to the conventional triangle
anomaly; the coefficient is half the conventional one. We recall that the
coefficient of the conventional triangle anomaly is fixed by imposing the
conservation law of the vector currents. Being consistent with this fact, we
observe that the vector current is not conserved with~Eq.~\eqref{eq:(4.3)}:
\begin{equation}
   -ip_\nu\left.\mathcal{T}_{\mu\nu\rho}^{ABC}(p,q)\right|_{O(g^0)}
   =\frac{1}{32\pi^2}\tr_R(1)\tr\left(t^A\{t^B,t^C\}\right)
   \epsilon_{\alpha\mu\beta\rho}p_\alpha q_\beta.   
\label{eq:(4.6)}
\end{equation}
Do these observations imply that our universal formula~\eqref{eq:(1.7)} is
incompatible with the physical requirement that the vector currents are
conserved (i.e., vector gauge invariance)?

We should note, however, that it is not a priori clear whether the small
flow-time expansion~\eqref{eq:(3.2)} for~$t\to0$ holds even if the point~$x$
collides with other composite operators in position space (see the discussion
in~Sect.~4.1 of~Ref.~\cite{Makino:2014taa}, for example). The
integration~\eqref{eq:(4.1)} in fact contains the correlation function at equal
points, $x=0$ or~$y=0$. Therefore, there exists freedom to modify
Eq.~\eqref{eq:(4.3)} by adding a term that contributes only when $x=0$ or~$y=0$
in position space. Using this freedom, we can redefine the correlation
function as
\begin{align}
   &\left\langle j_{5\mu}^A(0)j_\nu^B(x)j_\rho^C(y)\right\rangle
\notag\\
   &\to\left\langle j_{5\mu}^A(0)j_\nu^B(x)j_\rho^C(y)\right\rangle
   -\frac{1}{16\pi^2}\tr_R(1)\tr\left(t^A\{t^B,t^C\}\right)
   \epsilon_{\mu\nu\rho\sigma}
   \left[\partial_\sigma\delta^4(x)\delta^4(y)
   -\delta^4(x)\partial_\sigma\delta(y)\right].
\label{eq:(4.7)}
\end{align}
This redefinition preserves the Bose symmetry among the vector currents and,
of course, does not affect the correlation function when $x\neq0$ or~$y\neq0$.

After the redefinition, we have
\begin{align}
   &i(p+q)_\mu\left.\mathcal{T}_{\mu\nu\rho}^{ABC}(p,q)\right|_{O(g^0)}
   =-\frac{1}{8\pi^2}\tr_R(1)\tr\left(t^A\{t^B,t^C\}\right)
   \epsilon_{\alpha\nu\beta\rho}p_\alpha q_\beta,
\label{eq:(4.8)}\\
   &-ip_\nu\left.\mathcal{T}_{\mu\nu\rho}^{ABC}(p,q)\right|_{O(g^0)}
   =-iq_\rho\left.\mathcal{T}_{\mu\nu\rho}^{ABC}(p,q)\right|_{O(g^0)}
   =0.
\label{eq:(4.9)}
\end{align}
These expressions coincide with the conventional form of the triangle anomaly.
Since what we added in~Eq.~\eqref{eq:(4.7)} is simply a term that vanishes
for~$x\neq0$ or~$y\neq0$, our computation above shows that the universal
formula~\eqref{eq:(1.7)} produces non-local structure that is consistent with
the triangle anomaly, at least in the lowest non-trivial order of perturbation
theory.

\section{Conclusion}
\label{sec:5}
In this paper, we presented a universal formula that expresses a
correctly normalized flavor non-singlet axial-vector current through the
gradient/Wilson flow. The formula is universal in the sense that it holds
irrespective of the chosen regularization, and especially holds with lattice
regularization. Whether our formula possesses possible advantages over past
methods in actual lattice Monte Carlo simulations must still be carefully
examined. In particular, the small~$t$ limit in~Eq.~\eqref{eq:(1.7)} is limited
by the lattice spacing as~$a\ll\sqrt{8t}$ and there exists a systematic error
associated with the extrapolation for~$t\to0$. See
Refs.~\cite{Asakawa:2013laa} and~\cite{Kitazawa:2014uxa} for the $t\to0$
extrapolation in Monte Carlo simulations with the universal formula for the
energy--momentum tensor.

As a purely theoretical aspect of our formula, it is interesting to note that
if one can show that the triangle anomaly obtained
by~$i(p+q)_\mu\mathcal{T}_{\mu\nu\rho}^{ABC}(p,q)$ from~Eq.~\eqref{eq:(4.1)} is
local in the sense that it is a \emph{polynomial\/} of $p$ and~$q$ (that we
believe is quite possible), then we may repeat the proof~\cite{Zee:1972zt} of
the Adler--Bardeen theorem~\cite{Adler:1969er}, i.e., the triangle anomaly does
not receive any correction by strong interaction.

It must also be interesting to generalize the construction in the present paper
to the flavor singlet axial-vector current. Here, one has to incorporate the
mixing with the topological charge density. Then it is quite conceivable that
this construction give a further insight on the nature of the topological
susceptibility defined through the gradient/Wilson
flow~\cite{Luscher:2010iy,Noaki:2014ura}.

\section*{Acknowledgments}

Participation at the KEK lattice gauge theory school organized by JICFuS was
quite helpful to us and we would like to thank the organizers for their
hospitality.
The work of H.~S. is supported in part by Grant-in-Aid for Scientific
Research~23540330.

\end{document}